\documentclass[aps,pre,amsmath,lengthcheck,superscriptaddress]{revtex4-2}

\usepackage{graphicx}\graphicspath{ {figures/} }
\usepackage{hyperref}
\hypersetup{colorlinks,allcolors=blue,breaklinks}

\newcommand{\be}{\begin{equation}}
\newcommand{\ee}{\end{equation}}
\newcommand{\ba}{\begin{align}}
\newcommand{\ea}{\end{align}}
\newcommand{\bi}{\begin{itemize}}
\newcommand{\ei}{\end{itemize}}

\newcommand{\la}{\left\langle}
\newcommand{\ra}{\right\rangle}
\newcommand{\pd}{\partial}

\newcommand{\bla}{bla\\bla\\bla\\bla\\bla}

\newcommand{\mb}[1]{\mbox{\boldmath$#1$}}
\newcommand{\mc}[1]{\mathcal{#1}}

\begin{document}

\title{Casimir-Onsager matrix for weakly driven processes}

\author{Pierre Naz\'e}
\email{pierrenaze@gmail.com}

\affiliation{\it Universidade Federal do Par\'a, Faculdade de F\'isica, ICEN,
Av. Augusto Correa, 1, Guam\'a, 66075-110, Bel\'em, Par\'a, Brazil}

\date{\today}

\begin{abstract}

Modeling of physical systems must be based on their suitability to unavoidable physical laws. In this work, in the context of classical, isothermal, finite-time, and weak drivings, I demonstrate that physical systems, driven simultaneously at the same rate in two or more external parameters, must have the Fourier transform of their relaxation functions composing a positive-definite matrix to satisfy the Second Law of Thermodynamics. By evaluating them in the limit of near-to-equilibrium processes, I identify that such coefficients are the Casimir-Onsager ones. The result is verified in paradigmatic models of the overdamped and underdamped white noise Brownian motions. Finally, an extension to thermally isolated systems is made by using the time-averaged Casimir-Onsager matrix, in which the example of the harmonic oscillator is presented. 

\end{abstract}

\maketitle

\section{Introduction}
\label{sec:intro}

Modeling a physical system can be very complicated if one is not prepared to choose the model properly. In this sense, unavoidable physical laws that the system should respect are on our side to help us in this unfortunate task. In Ref.~\cite{naze2020}, my co-author and I have provided criteria for linear-response theory to be compatible with the Second Law of Thermodynamics in the finite-time and weak regime processes: the relaxation function that models the system, which can be chosen phenomenologically, has to present its Fourier transform positive.  However, this was made for driven systems performed by the perturbation of one single parameter. In this work, I provide an extension to such criteria, considering drivings at the same rate with two or more external parameters in the same context. 

Such extension is based again on the application of Bochner's theorem for a set of external parameters~\cite{feller1991introduction}. In this case, the Fourier transforms of the relaxation functions of the system must compose a positive-definite matrix. Given the similarity with the Casimir-Onsager matrix on such property~\cite{onsager1931reciprocal,casimir1945}, I investigated the behavior of this new matrix in the near-to-equilibrium regime, where the rate of the process is very large with respect to the relaxation time of the system. I verified then that this new matrix reduces to the Casimir-Onsager one in this case. Therefore, an extension of such a concept to finite-time and weak drivings can be considered. 

The positive definiteness of a particular matrix in the context of weak drivings already appears in the literature. I remark for example that such property is implicit in considerations of energy dissipation of systems performing weak drivings in Kubo's textbook~\cite{kubo2012}. In particular, Andrieux and Gaspard, in Ref.~\cite{gaspard2008}, claim that they did a general extension of Casimir-Onsager coefficients by using linear-response theory. However, in my opinion, this extension was made incompletely. Indeed, their work only shows that the coefficients of conductivity tensor of finite-time and weak drivings are symmetric due to the time-reversal symmetry of the Hamiltonian of interest. Also, it does not show that their extension reduces to Casimir-Onsager reciprocity relations in the near-to-equilibrium regime. The present work aims to complete this extension.

Other extensions of Casimir-Onsager reciprocity coefficients have been made in the literature: to nonequilibrium stationary states and nonlocality~\cite{dufty1987}, to inhomogeneous media~\cite{van1991onsager}, chemical reactions~\cite{gaspard2004}, to open bosonic Gaussian states~\cite{salazar2020nonlinear}, to cross phenomena theory~\cite{liu2022theory}, and asymmetric coefficients~\cite{benenti2011,lee2020exactly}; applications has been made as well in the fields of chemical kinetics~\cite{grmela2018}, thermoelectrics~\cite{bell2008} and heat machines~\cite{izumida2021hierarchical}. 

To illustrate the consistency of our result, I analyze two paradigmatic examples that are theoretically well-studied and experimentally verified: the overdamped and underdamped white noise Brownian motion~\cite{seifert2012stochastic,tome2015stochastic}. I calculate then the Casimir-Onsager matrices related to the drivings at the same rates of the stiffening and moving laser traps and verify that they are indeed positive-definite.

Finally, a final extension is made considering thermally isolated systems, where the heat bath is removed when the external parameters change. After discussing the main concerns of doing this, I present the time-averaged approach of the work presented in previous works to overcome the problems~\cite{naze2023adiabatic,naze2023quantum}. By measuring the energy variation of the system with this new observable, the natural timescale of the system is accessible, allowing an extension to the Casimir-Onsager coefficients where the approximation to near-to-equilibrium is doable. The calculation of the time-averaged Casimir-Onsager matrix of the harmonic oscillator is presented at the end as an example.

\section{Linear-response theory}
\label{sec:lrt}

I start by defining notations and developing the main concepts used in this work. This section is based on the technical introductory section of Ref.~\cite{naze2022optimal}.

Consider a classical system with a Hamiltonian $\mc{H}(\mb{z}(\mb{z_0},t)),\lambda(t))$, where $\mb{z}(\mb{z_0},t)$ is a point in the phase space $\Gamma$ evolved from the initial point $\mb{z_0}$ until time $t$, with $\lambda(t)$ being a time-dependent external parameter. Initially, for being the easiest way to prepare it, the system is at equilibrium with heat bath of temperature $\beta\equiv {(k_B T)}^{-1}$, where $k_B$ is Boltzmann's constant. During a switching time $\tau$, the external parameter is changed from $\lambda_0$ to $\lambda_0+\delta\lambda$, with the system still being in contact with the heat bath. The average work performed on the system during this interval of time is
\be
\overline{W}(\tau) \equiv \int_0^\tau \la\overline{\pd_{\lambda}\mc{H}}(t)\ra_0\dot{\lambda}(t)dt,
\label{eq:work}
\ee
where $\partial_\lambda$ is the partial derivative in respect to $\lambda$ and the superscripted dot the total time derivative. The generalized force $\la\overline{\pd_{\lambda}\mc{H}}\ra_0$ is calculated using the averaging $\overline{\cdot}$ over the stochastic path and the averaging $\langle\cdot\rangle_0$ over the initial canonical ensemble. The external parameter can be expressed as
\be
\lambda(t) = \lambda_0+g(t)\delta\lambda,
\label{eq:ExternalParameter}
\ee
where to satisfy the initial conditions of the external parameter the protocol $g(t)$ must satisfy the following boundary conditions $g(0)=0$, $g(\tau)=1$.

Linear-response theory aims to express average quantities until the first-order of some perturbation parameter considering how this perturbation affects the observable to be averaged and the probabilistic distribution \cite{kubo2012}. In our case, we consider that the parameter does not considerably change during the process, $|g(t)\delta\lambda/\lambda_0|\ll 1$, for all $t\in[0,\tau]$ and $\lambda_0\neq 0$. The generalized force can be approximated until the first order as~\cite{naze2020}
\begin{equation}
\begin{split}
\la\overline{\pd_{\lambda}\mc{H}}(t)\ra_0 =&\, \la\pd_{\lambda}\mc{H}\ra_0-\widetilde{\Psi}_0 \lambda(t)\\
&+\int_0^t \Psi_0(t-t')\dot{\lambda}(t')dt',
\label{eq:genforce-relax}
\end{split}
\end{equation}
where 
\be
\Psi_0(t) = \beta\la\pd_\lambda\mc{H}(0)\overline{\pd_\lambda\mc{H}}(t)\ra_0-\mc{C}
\ee 
is the relaxation function and $\widetilde{\Psi}_0\equiv \Psi_0(0)-\la\pd_{\lambda\lambda}^2\mc{H}\ra_0$ \cite{kubo2012}. The constant $\mc{C}$ is calculated to vanish the relaxation function for long times \cite{kubo2012}. The relaxation time of the system is defined as~\cite{naze2020}
\be
\tau_R=\int_0^\infty \frac{\Psi_0(t)}{\Psi_0(0)}dt.
\ee
Combining Eqs.~\eqref{eq:work} and \eqref{eq:genforce-relax}, the average work performed at the linear response of the generalized force is
\begin{equation}
\begin{split}
\overline{W}(\tau) = &\, \delta\lambda\la\pd_{\lambda}\mc{H}\ra_0-\frac{\delta\lambda^2}{2}\widetilde{\Psi}_0\\
&+\int_0^\tau\int_0^t \Psi_0(t-t')\dot{\lambda}(t')\dot{\lambda}(t)dt'dt.
\label{eq:work2}
\end{split}
\end{equation}

\begin{figure}[h]
    \centering
    \includegraphics[scale=0.35]{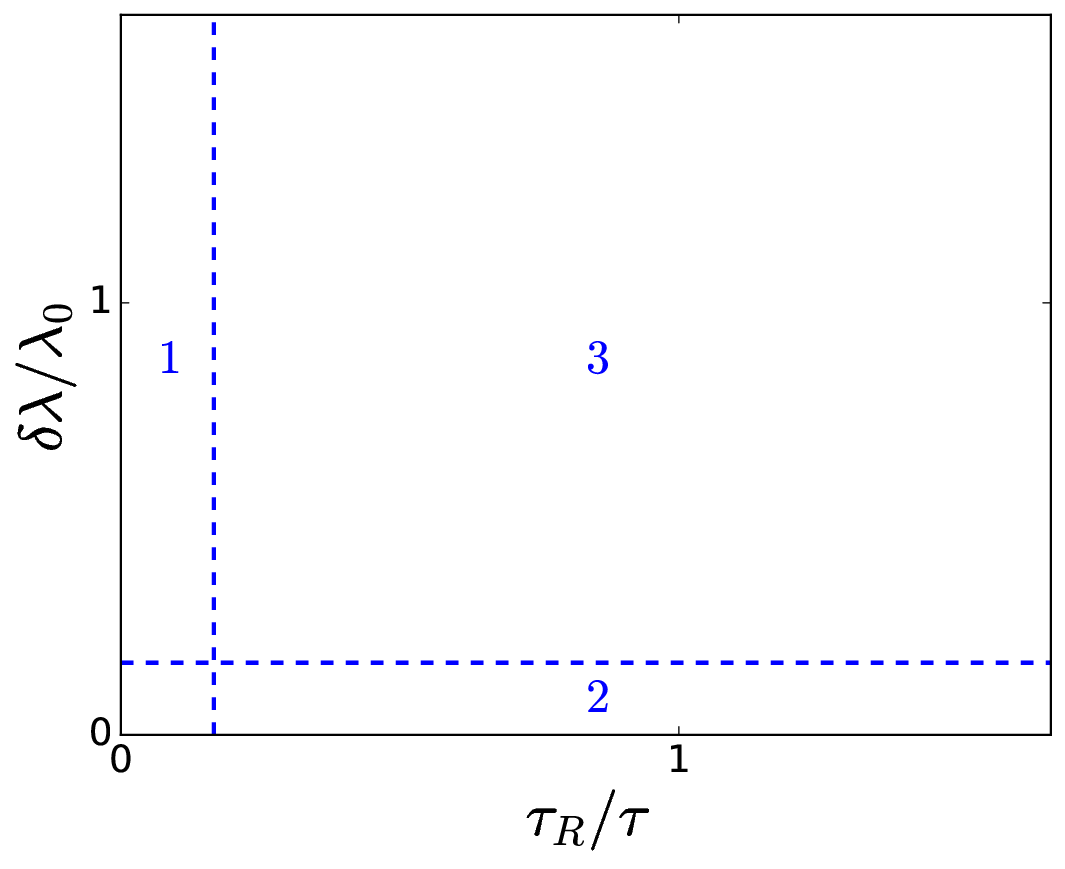}
    \caption{Diagram of non-equilibrium regimes. In region 1, the system is at near-to-equilibrium regime ($\tau\gg\tau_R$), while in region 2 on weakly driven processes ($\delta\lambda/\lambda_0\ll 1$). Region 3 is arbitrarily out-of-equilibrium.}
    \label{fig:1}
\end{figure}

We observe that the double integral on Eq.~\eqref{eq:work2} vanishes for long switching times \cite{naze2020}, which indicates that the other terms are the contribution of the difference of Helmholtz's free energy. The irreversible work $W_{\rm irr}$ is therefore
\begin{equation}
\begin{split}
W_{\rm irr}(\tau) = \frac{1}{2} \int_0^\tau\int_0^\tau \Psi_0(t-t')\dot{\lambda}(t')\dot{\lambda}(t)dt'dt,
\label{eq:wirrLR}
\end{split}
\end{equation}
where the symmetric property of the relaxation function was used \cite{kubo2012}. Such expression is valid for what is called finite-time and weak regime, where $\delta\lambda/\lambda_0\ll 1$. On the other hand, in near-to-equilibrium and weakly driven regime, where $\tau\gg\tau_R$ and $\delta\lambda/\lambda_0\ll 1$, one can approximate the relaxation function to~\cite{naze2022optimal}
\be
\lim_{\tau/\tau_R\gg 1}\Psi_0(t)\approx 2\tau_R \Psi_0(0)\delta(t),
\ee
such that the irreversible work becomes
\be
W_{\rm irr}(\tau)=\tau_R\Psi_0(0)\int_0^\tau\dot{\lambda}^2(t)dt.
\ee
In Fig.~\ref{fig:1}, the regimes treated in this work are depicted according to their dependence between $\delta\lambda/\lambda_0$ and $\tau_R/\tau$.

It is interesting to observe the behavior of the generalized force in near-to-equilibrium and weakly driven processes. In this case, one has
\be
\langle\overline{\partial_{\lambda} \mathcal{H}}(t)\rangle_0 = \la\pd_{\lambda}\mc{H}\ra_0-\widetilde{\Psi}_0 \lambda(t)+2\tau_R\Psi_0(0)\dot{\lambda}(t),
\ee
where the term $2\tau_R\Psi_0(0)$ connects the generalized force linearly with the flux $\dot{\lambda}(t)$~\cite{naze2020}. This is expected since, in this regime, the system does not present any ``memory'' and must depend instantly on time. This is the definition of the usual Casimir-Onsager reciprocity coefficient~\cite{callen1998}. Indeed, Irreversible Linear Thermodynamics, where those coefficients were originally defined, deals with phenomena in region 1 (see Fig.~\ref{fig:1}).

The objective of this work is to find conditions to satisfy the Second Law of Thermodynamics, that is, $W_{\rm irr}\ge 0$, when the physical system is driven simultaneously at the same rate by two or more external parameters. This proceeding will lead to the extension of the Casimir-Onsager matrix to finite-time and weak processes.

\section{Extended Casimir-Onsager matrix}
\label{sec:onsager}

Consider now that the Hamiltonian of interest presents a $n$ number of external parameters $\lambda_l(t)$, which are simultaneously driven at the same rate. In this case, the average work will be
\be
\overline{W}(\tau) \equiv \sum_{l=1}^n\int_0^\tau \la\overline{\pd_{\lambda_l}\mc{H}}(t)\ra_0\dot{\lambda_l}(t)dt,
\label{eq:work3}
\ee
Proceeding in the same fashion as we have done in Sec.~\ref{sec:lrt}, the irreversible work in the linear-response regime will be
\begin{equation}
W_{\rm irr}(\tau) = \frac{1}{2}\sum_{l,m=1}^n\int_0^\tau\int_0^\tau \Psi_{l m}(t-t')\dot{\lambda_l}(t')\dot{\lambda_m}(t)dt'dt,
\label{eq:wirr2}
\end{equation}
where
\be
\Psi_{l m}(t) = \beta\langle\partial_l \mathcal{H}(0)\partial_m \mathcal{H}(t)\rangle_0-\mathcal{C}_{l m}.
\label{eq:psi2}
\ee
We remark that the even symmetry used to achieve Eq. \eqref{eq:wirr2} occur for $\Psi_{ll}(t)$ and $\Psi_{l m}(t)+\Psi_{m l}(t)$, $l\neq m$. Also, $C_{l m}$ are chosen to nullify the relaxation functions for long times. Analogously to what was done in Ref.~\cite{naze2020}, I demand as a basic condition that the Laplace transforms of the relaxation functions will be finite on $s=0$. We can then define the relaxation time for each relaxation function
\be
\tau_R^{l m}=\int_0^\infty \frac{\Psi_{l m}(t)}{\Psi_{l m}(0)}dt.
\ee
To obtain $W_{\rm irr}\ge 0$, we must have the following matrix
\be
\mathcal{O} = [\hat{\Psi}_{l m}(\omega)]
\label{eq:casimironsagermatrix}
\ee
positive-definite for any $\omega\in\mathcal{R}$. For an explicit deduction see Appendix \ref{app:A}. Here, $\hat{\Psi}_0(\omega)$ is the Fourier transform of $\Psi_0(t)$
\be
\hat{\Psi}_0(\omega)=\frac{1}{\sqrt{2\pi}}\int_{\mathcal{R}}\Psi_0(t)e^{-i\omega t}dt.
\ee
The matrix $\mathcal{O}$ being positive-definite means that $\mathcal{O}$ should have all its eigenvalues non-negative. Also, for instance, for $2\times 2$ matrices, it is enough to show that $\mathcal{O}$ is symmetric and
\be
{\rm det}\,\mathcal{O}\ge 0,\quad {\rm Tr}\,\mathcal{O}\ge 0,
\ee
where ${\rm det}$ and ${\rm Tr}$ are respectively its determinant and trace. In particular, given the time-reversal symmetric property $\Psi_{l m}(t)=\Psi_{m l}(-t)$, it holds $\hat{\Psi}_{l m}(\omega)=\hat{\Psi}_{ml}(-\omega)$. This means that when the Fourier transform does not depend on $\omega$ the coefficients with reverse indexes are identical.

However, the question remains: what is the meaning of the coefficients $\hat{\Psi}_{l m}(\omega)$? I claim that they are the generalization of Casimir-Onsager reciprocity coefficients for finite-time and weak processes. To see that, consider the near-to-equilibrium and weak regime where Casimir-Onsager reciprocity coefficients are defined. In this case, it holds the approximation~\cite{naze2022optimal}
\be
\lim_{\tau/\tau_R^{lm} \gg 1}\Psi_{l m}(t)\approx 2\tau_R^{l m}\Psi_{l m}(0)\delta(t).
\ee
Their Fourier transforms are
\be
\lim_{\tau/\tau_R^{lm} \gg 1}\hat{\Psi}_{l m}=\sqrt{\frac{2}{\pi}}\tau_R^{l m}\Psi_{l m}(0).
\ee
Therefore, it holds
\be
\lim_{\tau/\tau_R^{lm} \gg 1}\Psi_{l m}(t-t')=\sqrt{2\pi}\left[\lim_{\tau/\tau_R^{lm} \gg 1}\hat{\Psi}_{l m}\right]\delta(t-t').
\label{eq:slowly}
\ee
The expression for the generalized force is
\be
\langle\overline{\partial_{\lambda_l} \mathcal{H}}(t)\rangle_0 = \sum_{m=1}^n\int_0^t \Psi_{l m}(t-t')\dot{\lambda}_m (t')dt',
\label{eq:genforce2}
\ee
where I have omitted the part with instantaneous memory. Applying Eq.~\eqref{eq:slowly} in Eq.~\eqref{eq:genforce2}, we have
\be
\langle\overline{\partial_{\lambda_l} \mathcal{H}}(t)\rangle_0 = \sqrt{2\pi}\sum_{m=1}^n\left[\lim_{\tau/\tau_R^{lm} \gg 1}\hat{\Psi}_{l m}\right]\dot{\lambda}_m (t),
\ee
which means that the coefficients of the matrix $\mathcal{O}$ are the Casimir-Onsager reciprocity ones in the near-to-equilibrium regime, where the fluxes are given by $\dot{\lambda}_m(t)$~\cite{naze2020}. Observe that such coefficients do not depend on $\omega$, meaning that the coefficients with reverse indexes are equal, as are the usual Casimir-Onsager reciprocity coefficients~\cite{callen1998}. This equality implies that the interplay between the variation of a parameter and the generalized force of another one is made equally if the parameters are reversed.

Returning to finite-time processes, the role that the extended Casimir-Onsager reciprocity coefficients play between the generalized force and the perturbation is rather complicated. Indeed
\be
\langle\overline{\partial_{\lambda_l} \mathcal{H}}(t)\rangle_0 = \sum_{m=1}^n\frac{1}{\sqrt{2\pi}}\int_{\mathcal{R}}\hat{\Psi}_{l m}(\omega)(e^{i \omega t'}\ast \dot{\lambda}(t'))(t)d\omega,
\ee
where
\be
(e^{i \omega t'}\ast \dot{\lambda}(t'))(t) = \int_0^t e^{i\omega(t-t')}\dot{\lambda}(t')dt'.
\ee
Instead, a better connection between both quantities is made by the Laplace transform of the relaxation functions
\be
\langle\widetilde{\overline{\partial_{\lambda_l} \mathcal{H}}}(s)\rangle_0 = \sum_{m=1}^n \widetilde{\Psi}_{l m}(s)\widetilde{\dot{\lambda}}_m (s),
\ee
where
\be
\widetilde{f}(s)=\int_0^\infty f(t)e^{-s t}dt.
\ee

\section{Examples: Brownian motion}
\label{sec:example}

I shall illustrate the consistency of our result showing the positive-definiteness of the matrices $\mathcal{O}$ for the paradigmatic examples of an overdamped and underdamped white noise Brownian motion, driven simultaneously by the stiffening and moving laser traps~\cite{tome2015stochastic,seifert2012stochastic}. 

\subsection{Overdamped case}

I consider first an overdamped Brownian particle, whose dynamics are governed by the following Langevin equation
\be
\dot{x}(t)+\frac{1}{\gamma}\pd_x\mc{V}(x(t),\lambda(t),\mu(t)) =\eta(t),
\label{eq:langevin}
\ee  
where $x(t)$ is its position at the instant $t$, $\gamma$ is the damping coefficient, $\lambda(t)$ is the first control parameter, $\mu(t)$ the second one and $\eta(t)$ is a Gaussian white noise characterized by
\be
\overline{\eta(t)}=0, \quad \overline{\eta(t)\eta(t')}=\frac{2}{\gamma\beta}\delta(t-t').
\label{eq:bceta}
\ee
The time-dependent potential will be a stiffening and moving laser traps
\be
\mc{V}(x(t),\lambda(t),\mu(t))=\frac{\lambda(t)}{2}(x(t)-\mu(t))^2.
\label{eq:MovingLaserTrap}
\ee
In this case, its extended Casimir-Onsager matrix is
\be
\mathcal{O}=
\begin{bmatrix}
\sqrt{\frac{2}{\pi}}\frac{\gamma}{\beta \lambda_0}\frac{1}{4\lambda_0^2+\gamma^2\omega^2} & 0 \\
0 & \sqrt{\frac{2}{\pi}}\frac{\gamma}{\beta }\frac{1}{\lambda_0^2+\gamma^2\omega^2} 
\end{bmatrix},
\ee
which is positive-definite. Therefore, as expected, the overdamped white noise Brownian motion, when subjected to simultaneous driving in the stiffening and moving laser traps, obeys the Second Law of Thermodynamics, showing its consistency as a paradigmatic model. I remark that the terms of the second diagonal being null indicate that there is no equilibrium correlation between the generalized forces of both parameters. 

\subsection{Underdamped case}

I consider now an underdamped Brownian particle, whose dynamics are governed by the following Langevin equation
\be
\frac{m}{\gamma}\ddot{x}(t)+m\dot{x}(t)+\frac{1}{\gamma}\pd_x\mc{V}(x(t),\lambda(t),\mu(t)) =\eta(t),
\label{eq:langevin}
\ee  
where $m$ is its mass, $x(t)$ is its position at the instant $t$, $\gamma$ is the damping coefficient, $\lambda(t)$ is the first control parameter, $\mu(t)$ the second one and $\eta(t)$ is a Gaussian white noise characterized by
\be
\overline{\eta(t)}=0, \quad \overline{\eta(t)\eta(t')}=\frac{2m}{\gamma\beta}\delta(t-t').
\label{eq:bceta}
\ee
The time-dependent potential will be a stiffening and moving laser traps
\be
\mc{V}(x(t),\lambda(t),\mu(t))=\frac{m\lambda(t)}{2}(x(t)-\mu(t))^2,
\label{eq:MovingLaserTrap}
\ee
where $4\lambda_0>\gamma^2$. Its extended Casimir-Onsager matrix is
\be
\mathcal{O}=
\begin{bmatrix}
\hat{\Psi}_{\lambda\lambda}(\omega) & 0 \\
0 & \hat{\Psi}_{\mu\mu}(\omega) 
\end{bmatrix},
\ee
where
\be
\hat{\Psi}_{\lambda\lambda}(\omega)=\frac{\sqrt{\frac{2}{\pi }} \gamma  \left(\gamma ^4+\gamma ^2 \left(\lambda_0 +\omega_0^2\right)+\lambda_0 
   \left(\omega ^2+\omega_0^2\right)\right)}{\beta  \lambda_0 ^2 \left(\gamma ^2+\omega ^2\right)
   \left(\gamma ^4+2 \gamma ^2 \left(\omega ^2+\omega_0^2\right)+\left(\omega ^2-\omega_0^2\right)^2\right)}
\ee
and
\be
\hat{\Psi}_{\mu\mu}(\omega)=\frac{4 m\sqrt{\frac{2}{\pi }} \gamma  \lambda_0 \left(\gamma ^2+\omega_0^2\right)}{\gamma ^4+2 \gamma
   ^2 \left(4 \omega ^2+\omega_0^2\right)+\left(\omega_0^2-4 \omega ^2\right)^2}
\ee
with $\omega_0=\sqrt{4\lambda_0-\gamma^2}>0$. Again, the extended Casimir-Onsager matrix $\mathcal{O}$ is positive-definite. Therefore, as expected, the underdamped white noise Brownian motion obeys the Second Law of Thermodynamics when subjected to simultaneous driving in the stiffening and moving laser traps, showing its consistency as a paradigmatic model. I remark that the terms of the second diagonal being null indicate no equilibrium correlation between the generalized forces of both parameters.

\section{Extension for thermally isolated system}

Until this point, I consider only isothermal drivings, where the system of interest is in contact always with the heat bath. An extension of the Casimir-Onsager matrix to thermally isolated systems, where the system is decoupled from the heat source during the change of the external parameters, requires a more careful analysis than this first case. Indeed, in a first attempt, the Casimir-Onsager matrix could be defined exactly by Eq.~\eqref{eq:casimironsagermatrix}. However, from this definition is not possible to extend such Casimir-Onsager coefficients in the near-to-equilibrium regime since such extensions are composed of the relaxation time of the system, which is not accessible by the theory proposed~\cite{naze2023adiabatic}.

To circumvent this problem, few works have presented an alternative theory to furnish such characteristic time of the system, and consequent extension to near-to-equilibrium regime~\cite{naze2023adiabatic,naze2023quantum}: to measure the variation of energy of the system, one uses instead the time-averaged irreversible work. In linear-response theory, this quantity is defined as
\be
\overline{W}_{\rm irr}=\frac{1}{2}\int_0^\tau \int_0^\tau \overline{\Psi}_0(t-t')\dot{\lambda}(t)\dot{\lambda}(t')dtdt',
\ee
where
\be
\overline{\Psi}_0(t)=\frac{1}{t}\int_0^t \Psi_0(u)du
\ee
is the time-averaged relaxation function. In this case, the decorrelation time of the thermally isolated system is
\be
\overline{\tau}_R=\int_0^\infty \frac{\overline{\Psi}_0(t)}{\overline{\Psi}_0(0)}dt,
\ee
which allows one to define the near-to-equilibrium regime approximation
\be
\lim_{\tau\gg\overline{\tau}_R}\overline{\Psi}_0(t)\approx 2\overline{\tau}_{R}\overline{\Psi}_0(0)\delta(t).
\ee
In this manner, for thermally isolated systems, a more suitable definition is the time-averaged Casimir-Onsager coefficients, given by the elements of the time-averaged Casimir-Onsager matrix
\be
\overline{\mathcal{O}}=[\hat{\overline{\Psi}}_{lm}(\omega)],
\ee
which is similar to the definition of the isothermal driving, but the relaxation function used is the time-averaged one.

Finally, as demonstrated in Ref.~\cite{naze2023adiabatic}, the positive-definiteness on the Casimir-Onsager matrix leads to the satisfaction of the Second Law of Thermodynamics by using equivalence between this law and the Second Law of Thermodynamics defined for the time-averaged case.

\section{Example: Classical harmonic oscillator}

To show the positive definiteness of the time-averaged Casimir-Onsager matrix, I calculate it for the paradigmatic case of the classical harmonic oscillator. Its Hamiltonian is
\be
\mathcal{H}=\frac{p^2}{2}+\frac{\lambda(t)}{2}(q(t)-\mu(t))^2,
\ee
where $\{q(t),p(t)\}$ are the canonical variables and $\lambda(t)$ and $\mu(t)$ are the time-dependent external parameters. The time-averaged Casimir-Onsager matrix is
\be
\overline{\mathcal{O}}=\begin{bmatrix}
\hat{\overline{\Psi}}_{\lambda\lambda}(\omega) & 0 \\
0 & \hat{\overline{\Psi}}_{\mu\mu}(\omega), 
\end{bmatrix}
\ee
where
\be
\hat{\overline{\Psi}}_{\lambda\lambda}(\omega)=\frac{1}{16}\frac{1}{\beta\lambda_0^{5/2}}\sqrt{\frac{\pi}{2}}(\text{sign}(2\sqrt{\lambda_0}-\omega)+\text{sign}(2\sqrt{\lambda_0}+\omega)),
\ee
and
\be
\hat{\overline{\Psi}}_{\mu\mu}(\omega)=\frac{\sqrt{\lambda_0}}{2}\sqrt{\frac{\pi}{2}}(\text{sign}(\sqrt{\lambda_0}-\omega)+\text{sign}(\sqrt{\lambda_0}+\omega)),
\ee
which are both non-negative. Therefore, the time-averaged Casimir-Onsager matrix is positive-definite. I remark that the terms of the second diagonal being null indicate that there is no equilibrium correlation between the generalized forces of both parameters.

\section{Final remarks}
\label{sec:final}

In this work, I have generalized the Casimir-Onsager matrix to finite-time and weak processes. I did so to extend the compatibility criteria of linear-response theory with the Second Law of Thermodynamics to driven processes performed at the same rate of two or more external parameters. The positive-definiteness of the extended Casimir-Onsager matrix is verified in two paradigmatic models of overdamped and underdamped white noise Brownian motions, driven simultaneously at the same rate in the stiffening and moving laser trap. An extension to thermally isolate systems is made at the end using the time-averaged Casimir-Onsager matrix. In all examples treated, the terms of the second diagonal are all null, indicating no equilibrium correlation between the generalized forces of different external parameters. I remark that the hypothesis of weak perturbation restricts potential application for more complex scenarios in finite-time processes, although a first step must be done to achieve this in further research. Finally, modeling the diffusive and thermoelectric phenomena of Irreversible Linear Thermodynamics in finite-time and weakly driven processes regime (region 2 in Fig.~\ref{fig:1}) in the Hamiltonian framework will be approached in future research.

\begin{acknowledgements}
    I acknowledge Jordan Horowitz and Akram Touil for enlightening discussions that led me to such a result some years ago. 
\end{acknowledgements}

\bibliography{OMLR}
\bibliographystyle{apsrev4-2}

\onecolumngrid

\appendix
\section{Second Law of Thermodynamics and Bochner's theorem}
\label{app:A}

Considering Eq.~\eqref{eq:wirr2}, we have
\begin{align}
W_{\rm irr}(\tau) &= \frac{\delta\lambda^2}{2}\sum_{l,m=1}^n\int_0^\tau\int_0^\tau \left(\frac{1}{\sqrt{2\pi}}\int_{\mathcal{R}}\hat{\Psi}_{l m}(\omega)e^{i \omega (t-t')} d\omega\right)\dot{\lambda_l}(t')\dot{\lambda_m}(t)dt'dt\\
&= \frac{\delta\lambda^2}{2\sqrt{2\pi}}\int_{\mathcal{R}}\left[\sum_{l,m=1}^n\hat{\Psi}_{l m}(\omega) \left(\int_0^\tau e^{i \omega t}\dot{\lambda_l}(t)dt\right)\left(\int_0^\tau e^{-i \omega t}\dot{\lambda_m}(t)dt\right)\right]d\omega\\
&=\frac{\delta\lambda^2}{2\sqrt{2\pi}}\int_{\mathcal{R}}\left[\sum_{l,m=1}^n\hat{\Psi}_{l m}(\omega) \alpha_l(\omega)\alpha_m(\omega)+\sum_{l,m=1}^n\hat{\Psi}_{l m}(\omega) \sigma_l(\omega)\sigma_m(\omega) \right]d\omega,\\
\end{align}
where
\be
\alpha_l(\omega)=\int_0^\tau \cos{\omega t}\dot{\lambda}_l(t)dt,\quad \sigma_l(\omega)=\int_0^\tau \sin{\omega t}\dot{\lambda}_l(t)dt.
\ee
Since the matrix $[\hat{\Psi}_{l m}]$ is positive-definite, accordingly to Bochner's theorem~\cite{feller1991introduction} it holds
\be
\sum_{l,m=1}^n\hat{\Psi}_{l m}(\omega) f_l(\omega)f_m(\omega)\ge 0,
\ee
for any set of continuous functions $\{f_l\}$. Therefore, $W_{\rm irr}(\tau)\ge 0$.
\end{document}